\documentclass[3p,times,twocolumn]{elsarticle}

\usepackage{amssymb}
\usepackage[figuresright]{rotating}

\begin{document}

\begin{frontmatter}



%% e.g. \dochead{17th International Conference on Dynamical Processes in Excited States of Solids}

\title{Search for correlations between solar flares and decay rate of radioactive nuclei}


\author[mi]{E. Bellotti}
\author[pd]{C. Broggini\corref{cor1}}
%\ead{broggini@pd.infn.it}
\author[lngs]{G. Di Carlo}
\author[lngs]{M. Laubenstein}
\author[pd]{R. Menegazzo}

\address[mi]{Universit\`{a} degli Studi di Milano Bicocca and Istituto Nazionale di Fisica Nucleare, Sezione di Milano, Milano, Italy}
\address[pd]{Istituto Nazionale di Fisica Nucleare, Sezione di Padova, Padova, Italy}
\address[lngs]{Istituto Nazionale di Fisica Nucleare, Laboratori Nazionali del Gran Sasso, Assergi (AQ), Italy}

\cortext[cor1]{Corresponding author}

\begin{abstract}
The deacay rate of three different radioactive
sources ($^{40}$K, $^{137}$Cs and $^{nat}$Th) has been measured with NaI and Ge detectors. Data have been analyzed to search for possible 
variations in coincidence with the two strongest solar flares 
of the years 2011 and 2012.  No significant deviations from standard expectation have been observed, with a few $10^{-4}$ sensitivity. As a consequence, we could not find
any effect like that recently reported by Jenkins and Fischbach: a few per mil decrease in the decay rate of $^{54}$Mn
during solar flares in December 2006.
\end{abstract}

\end{frontmatter}

% \linenumbers

\vspace*{-0.5cm}

\section{Introduction}
\label{}
In the past years, a correlation between the Sun and the decay rate of radioactive isotopes has 
been proposed. In particular, two effects have been considered: the annual modulation due 
to the seasonal variation of the Earth-Sun distance \cite{Jen09} and the decrease of the decay rate 
during a solar flare \cite{Jen06}. In this letter we are interested in  the latter phenomenon.
 
Briefly, solar flares are explosions on the surface of the Sun near sunspots. They are powered 
by the release of magnetic energy stored in the corona, up to one hundredth of the
solar luminosity, and they affect all layers of solar atmosphere, from the photosphere 
to the corona. On the Sun this amount of energy is released within a few minutes to tens 
of minutes. In this interval the plasma is heated to  tens of millions of degrees with a 
strong X-ray emission and electron and proton acceleration (up to several tens and hundreds of MeV, respectively).

In particular, the 2006 flares from December 2nd 2006 to January 1st 2007 gave rise to X-ray fluxes which, measured on the 
the Geostationary  Operational Environmental Satellites (GOES),
were of a few times 10$^{-4}$  W/m$^{2}$ at the peak (see Fig. 1 of reference
\cite{Jen06} for details).  
At that time the activity of a 
$\sim$1 $\mu$Ci source of $^{54}$Mn was being measured 
by Jenkins and Fischbach \cite{Jen06}
with a 2x2 inch NaI crystal detecting 
the 835 keV $\gamma$-ray emitted after the electron capture decay. 
A significant dip (up to 4$\cdot$10$^{-3}$, $\sim$7 $\sigma$ effect), in the count rate,
averaged on a time interval of 4 hours, has been observed in coincidence with the solar flares.
On the other hand, a different experiment with a $\sim$10$^{-3}$ sensitivity,
carried out by Parkhomov \cite{Par10}, did not observe any deviation in the 
activity of $^{60}$Co, $^{90}$Sr-Y and $^{239}$Pu sources in coincidence with the same flare.

After a few years of quiet Sun, solar activity is now increasing, as shown both by the 
increase of the steady X-ray flux as 
well as of X-flares and of other typical solar phenomena. As a matter of fact, we are 
approaching the maximum of the 11 year solar cycle which is predicted to take place in 
Fall 2013. In our analysis we focus on 
the two most intense flares of the last years, namely those that 
occurred on August  2011 and March 2012:
X6.9 on August 9th 2011 @ 08:08 UTC  and X5.4 on March 7th 2012 @ 00:24 UTC \cite{noaa}.
\begin{table*}[ht]
\caption{Experimental set-ups active at the time of large solar flares. The $^{137}$Cs and $^{nat}$Th measurements are running underground while the $^{40}$K set-up is installed in the external laboratory at Gran Sasso.}
\footnotesize
\begin{center}
\begin{tabular}{|c|c|c|c|c|c|c|}
\hline 
Flare time & Peak intensity & Source & detector & Observed decay type &
Mean counting rate & Integration time \\ 
\hline
August 9th 2011  08:08 UTC & X6.9 & $^{137}Cs$ & HPGe & $\beta^-$ & 680 Hz & 1 hour \\
March 7th 2012 00:24 UTC & X5.4 & & & & & \\
\hline
August 9th 2011  08:08 UTC & X6.9 & $^{nat}Th$ & NaI & $\alpha + \beta^-$ decay chain & 3300 Hz & 24 hours \\
\hline
March 7th 2012  00:24 UTC & X5.4 & $^{40}K$ & NaI & EC & 
790 Hz & 1 hour \\
\hline
\end{tabular}
\end{center}
\end{table*}

Solar flares are classified according to the power of the X-ray flux peak near the Earth as 
measured by the GOES-15 geostationary satellite:
X identifies the class of the most powerful ones, with a power at the peak larger than 10$^{-4}$  W/m$^{2}$ 
(within the X-class there is then a linear scale). 
The two flares were well defined in time (a few minutes) and they illuminated the entire 
Earth. Their intensities are comparable, or even larger, than those observed 
in December 2006. During the 2011 flare the activity of the $^{137}$Cs and $^{nat}$Th 
sources were being measured with a Ge and with a NaI detector, respectively. 
On the other hand, during the 2012 flare the $^{137}$Cs and $^{40}$K sources were being studied
with the same Ge detector and with a different NaI detector, as described in the next section.
These different nuclides gives the possibility to search for  possible
effects correlated with solar flares in three different decay processes: alpha, beta
and electron capture.

\section{The set-ups}
\label{}
Table 1 summarizes the information on the experimental set-ups we are running to search for
modulations in the decay rates of different radioactive sources (period
from few days to one year). In particular, in this letter we only consider an interval of $\pm 10$ days around 
the time of the 2 solar flares in order to search for any significant deviation (positive or negative) 
in the decay rate correlated with the flares.
The choice of this time window is quite arbitrary, since there is no 
model, to our knowledge, that correlates the flare intensity with the activity of a radioactive source. 
On the other hand, we note that, according to data shown in Fig. 2 of \cite{Jen06},
the alleged influence of the flare on the source activity lasts for a few days around 
the occurrence of the flare. 

\subsection{Potassium source}

A 3x3 inch NaI crystal is surrounded by about 16 kg of potassium bicarbonate powder. 
The set-up, installed above ground, is shielded by at least 10 cm of lead. 
The total count rate in the 17-3400 keV energy window is about 800 Hz, to be compared
to the background of less than 3 Hz when the source is removed. 
The energy spectrum is dominated by the full 
energy peak at 1461 keV energy due to the electron-capture decay of $^{40}$K to $^{40}$Ar. 
The peak position and the energy resolution ($\simeq 90$ keV at 1461 keV) are fairly constant 
over months. 

\subsection{Cesium source}

The activity of a 3 kBq $^{137}$Cs source is being measured since June 2011.
The set-up is installed in the low background facility STELLA (SubTErranean Low Level Assay) 
located in the underground laboratories of Laboratori Nazionali del Gran Sasso (LNGS). 
The detector is a p-type High Purity Germanium (96$\%$ efficiency) with the source firmly 
fixed to its copper end-cap and it is surrounded by at least 5 cm of
copper followed by 25 cm of lead to suppress the laboratory gamma ray background. Finally, 
shielding and detector are housed in a polymethylmetacrilate box flushed 
with nitrogen at slight overpressure and which is working as an anti-radon shield. 
The total count rate above the 7 keV threshold is of 680 Hz. 
The intrinsic background, i.e shielded detector without Cs source, 
has been measured during a period of 70 days: thanks to the underground environment and 
to the detector shielding, it is very low, down to about 40 counts/hour above the 
threshold (0.01 Hz).
The spectrum is dominated by the 661.6 keV line due to the isomeric transition of
$^{137m}$Ba from the beta decay of $^{137}$Cs.

Details of the experiment and the results obtained in the first 210 days of running 
to search for an annual modulation of the $^{137}$Cs decay constant are given in \cite{Bel12}. 
Briefly, a limit of 8.5$\cdot$10$^{-5}$ at 95$\%$ C.L. is set
on the maximum allowed amplitude independently of the phase. 

\subsection{Thorium source}

The activity of a sample of natural Thorium is measured with a 3x3 inch NaI crystal 
installed underground in the same laboratory as the Germanium experiment with the $^{137}$Cs source. The sample is an optical lens, made by special glass heavily doped with Thorium Oxide. 
Note that this technique, used for improving the optical properties of glass, was quite common until 
the seventies. The lens is placed close to the crystal housing and both the lens and the NaI 
detector are shielded with at least 15 cm of lead. 
The total count rate above the threshold of 10 keV is of about 3200 Hz
(gammas from $^{228}$Ac , $^{212}$Bi, $^{212}$Pb, $^{208}$Tl), with a background of 2.3 Hz
(due to $^{40}$K, thorium and uranium chains and lead X-rays). 
The energy spectrum is acquired once a day, with a corrected dead time of 2.63$\%$.
Even if the chain is not at the equilibrium, the total count rate increases
by only $1.7\cdot10^{-4}$ over a time period of 1 month.

\begin{figure*}[htb] %  figure placement: here
\centering
\includegraphics[width=6.20in]{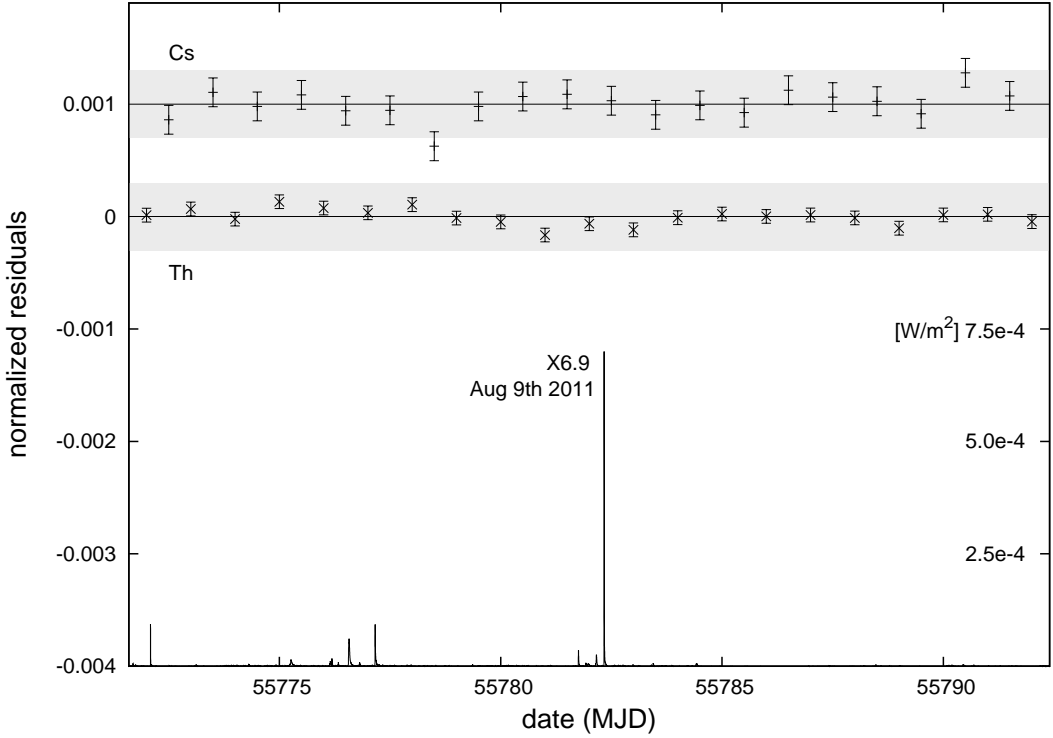} 
\vspace*{-0.5cm} % only if needed
\caption{Residuals, averaged over 1 day, of $^{137}$Cs and $^{nat}$Th data collected around the August 9th 2011 flare and the X-ray flux 
measured by the GOES-15 Satellite.
The $^{137}$Cs data are vertically displaced by 0.001 for sake of clarity; the right-hand vertical scale
gives the X-ray flux measured in W/m$^2$. The two shaded bands are drawn at $\pm 3\cdot 10^{-4}$
from the expected value.}
\label{fig:dati1}
\end{figure*}

\section{Results}

We consider separately the two largest solar flares occurred in the data taking period, 
{\it i.e.} X6.9 August 9th 2011 and X5.4 March 7th 2012 \cite{noaa}. For each of them only
two of the set-ups given in Table 1 were running. As a matter of fact, the $^{nat}$Th set-up went out of order in February 2012, due to a
failure in the DAQ system, whereas the $^{40}$K set-up started taking data in November 2011. On the contrary, the $^{137}$Cs set-up 
is continuously running since June 2011.

Figure 1 shows the data collected in a 20 day window centered on the August 9th 2011 flare
(the day is given in terms of the Modified Julian Date).
The X-ray peak flux is plotted in linear scale and given in W/m$^2$, in the 0.1-0.8 nm
band measured by the GOES-15 satellite \cite{noaa}. Inside the two bands 
are plotted the residuals
of the normalized count rate of the $^{nat}$Th  and $^{137}$Cs sources (i.e. the difference between the measured and expected count rate divided by the measured one), averaged over a period of 1 day. 

\begin{figure*}[htb] %  figure placement: here
\centering
\includegraphics[width=6.20in]{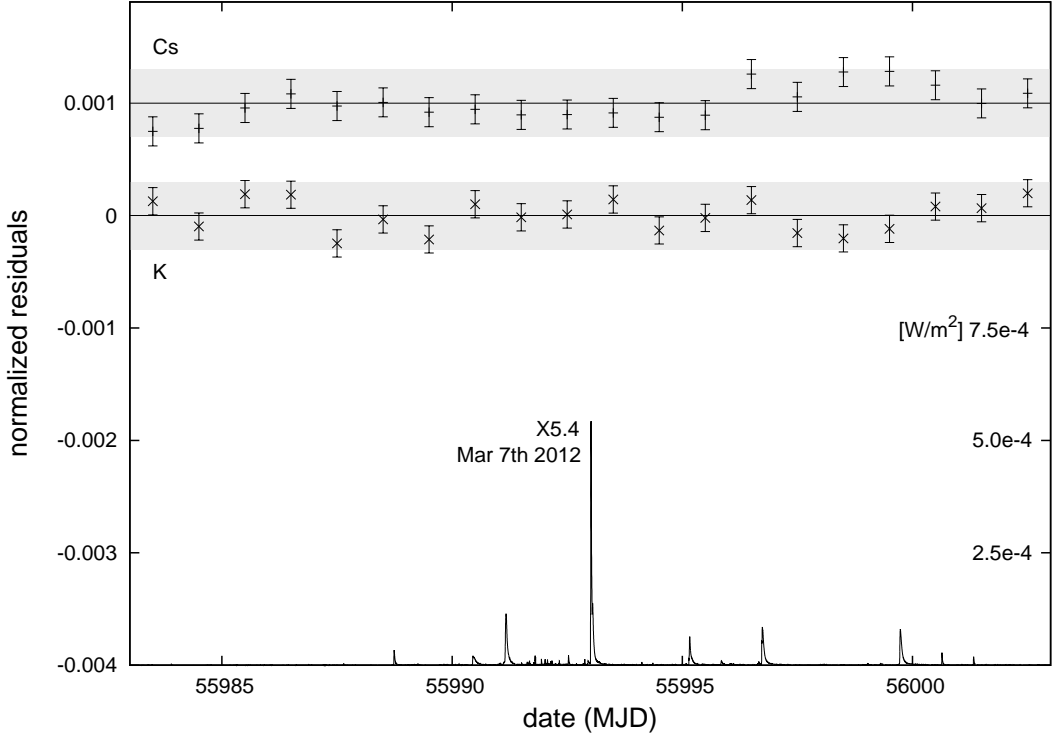} 
\vspace*{-0.5cm} % only if needed
\caption{Residuals, averaged over 1 day, of $^{137}$Cs and $^{40}$K data collected around the March 7th 2012 flare and the X-ray flux 
measured by the GOES-15 Satellite.
The $^{137}$Cs data are vertically displaced by 0.001 for sake of clarity; the right-hand vertical scale
gives the X-ray flux measured in W/m$^2$. The two shaded bands are drawn at $\pm 3\cdot 10^{-4}$
from the expected value.}
\label{fig:dati2}
\end{figure*}

The error bars are purely statistical. Systematic errors are negligible as compared to the statistical ones during a data taking period of 
a few days only.
For the $^{nat}$Th data a linear trend (5.7 ppm/day), due to the recovering of the secular equilibrium,
is subtracted, while the $^{137}$Cs data are corrected for the exponential decay of the source, using
the nominal mean life value of 43.38 y. This latter correction amounts to 63 ppm/day. 

From the data we can conclude that the $^{137}$Cs source does not show any significant dip or excess 
in correspondence with the X-ray main peak. On the other hand, the $^{nat}$Th source shows a questionable dip in the 
count rate, starting 1.5 days before the X-flare. However, the dip is well compatible with a statistical fluctuation.
As a matter of fact, fluctuations of the same order 
of magnitude can be seen at different times during the data taking, uncorrelated with X-ray flux peaks. 
In any case, the existence in our data of an effect as large as the one reported in \cite{Jen06}, of the order of a few per mil per day and lasting 
several days, can be excluded.
The maximum effect compatible with our data is smaller than $3\cdot 10^{-4}$ per day 
at $95\%$ confidence level for the X6.9 flare. Such a limit is obtained by adding the double of the error to the value of the dip.

In Figure 2 similar data for the March 7th 2012 flare are presented. Also in this case, no significant effect can be
seen related to the occurrence of the X5.4 flare, both in the $^{137}$Cs and in the $^{40}$K data.
An upper limit similar to the one given above can be issued by taking twice the statistical error. 
Note that no effect can be
seen also in correspondence with the arrival on earth of the two CME (coronal mass emission)
related to this earth facing flare, respectively on March 8th and 11th 2012 (55994 and 
55997 MJD). 

\begin{figure*}[htb] %  figure placement: here
   \centering
   \includegraphics[width=6.10in]{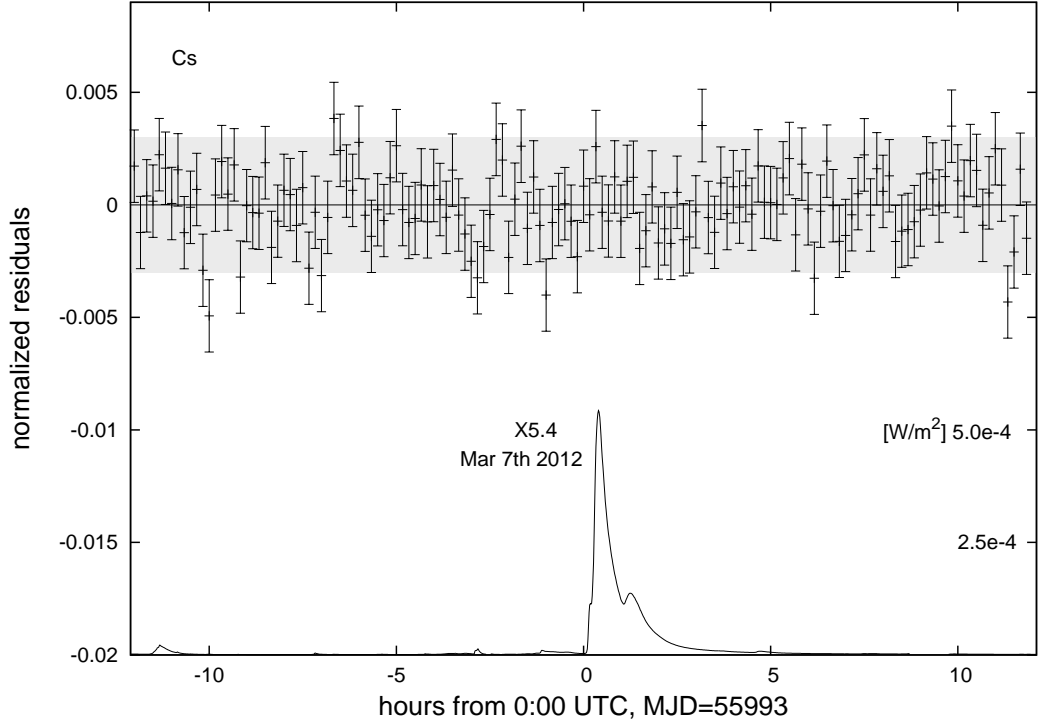} 
   \caption{Residuals of $^{137}$Cs data, averaged over 10 minutes, collected around March 7th 2012 flare and X-ray flux 
from the GOES-15 Satellite. The right-hand vertical scale
gives the X-ray flux measured in W/m$^2$. The shaded band is drawn at $\pm 3 \cdot 10^{-3}$
from the expected value.}
   \label{fig:dati3}
\end{figure*}

During this March 7th 2012 flare we were taking the $^{137}$Cs data also with a fully digital list mode data acquisition system, recording the
time of each event. As a consequence, we can have the source count rate averaged over shorter time than the day or the hour.
In particular, Figure 3 shows the  $^{137}$Cs source residuals
averaged over 10 minutes in a 24 hour time window containing the X-ray peak.
Again, no fast occurring effect incompatible with a statistical fluctuation and larger than $3\cdot 10^{-3}$ (at $95\% C.L.$) can be seen (we note that the sensitivity
scales linearly with the square root of the averaging time).
We cannot repeat this analysis for the 2011
flare because at the very timing of the flare the data acquisition
has been stopped for an hour due to liquid nitrogen refilling of the Germanium detector.

\section{Conclusion}

The gamma activity of three different sources, $^{40}$K (electron capture) ,  $^{137}$Cs 
(beta decay) and $^{nat}$Th (alpha and beta decays)  
have been measured during the occurrence of at least one of the two strongest solar 
flares of the years 2011 and 2012. No significant deviations from expectations have been observed.
Up to now there are no quantitative models able to correlate the flare intensity with the 
decay constant of radioactive isotopes. However, from our data 
it is possible to conclude that a universal deviation of decay rate 
(alpha or beta or electron capture decay) is less than $3\cdot$10$^{-4}$ per day 
for a flare of $7\cdot$10$^{-4}$ W/m$^{2}$ flux at the peak. By 'universal' we mean a deviation in the 
count rate affecting in the same way all radioactive isotopes decaying through the same basic mechanism.
We are now continuing the life-time measurements to search for modulations in the decay rate
of different nuclides. This way we will have the opportunity to further investigate the issue 
of the solar flare correlation with nuclear decay in case stronger flares 
should happen closer in time to the expected 2013 solar maximum.

\section {Acknowledgments}
The Director of the LNGS and the staff of the Laboratory are warmly acknowledged for their 
support. We want to thank also Prof. Roberto Battiston for his continuous encouragement.



%% References


%\bibliographystyle{elsarticle-num}



\end{document}